\documentclass{ws-ijmpe}

\usepackage{psfrag}

\begin{document}

\markboth{C.~Hanhart}{Non-Strange Meson Production with Hadronic
Probes}

\catchline{}{}{}{}{}
%%%%%%%%%%%%%%%%%%%%%%%%%%%%%%%%%%%%%%%%%%%%%%%%%%%%%%%%%%%%%%%%%%%%

\title{NON--STRANGE MESON PRODUCTION\\ WITH HADRONIC PROBES}

\author{\footnotesize C.~HANHART}

\address{Institut f\"ur Kernphysik and J\"ulich Center for Hadron Physics, \\
Forschungzentrum J\"ulich, D-52425 J\"ulich, Germany \\
c.hanhart@fz-juelich.de}

\maketitle

\begin{history}
\received{11 September 2008} \revised{24 September 2008}
%\accepted{(Day Month Year)}
%\comby{(xxxxxxxxxx)}
\end{history}

% for author index, please do not change
\index{Hanhart C.}

\begin{abstract}
Recent progress in the field of meson production in hadron
collisions is presented. Special emphasis is put new developments in
the application of effective  field theory methods to pion
production, but also theoretical  concepts for two--pion, $\eta$,
and vector--meson production will be  presented.
\end{abstract}

\section{Introduction}

As a consequence of the large coupling strength of Quantum--Chromo
Dynamics (QCD) at low and intermediate energies, standard
perturbative methods fail --- in fact, the interaction gets so
strong at increasing distances that it confines all color--charged
particles (quarks and gluons) into areas of the order of $10^{-15}$
m --- only those particles that are neutral with respect to the
color force (these are called hadrons) are allowed to travel longer
distances. Due to this complication, strong QCD is one of the last
\emph{terrae incognitae} of the Standard Model.

To theoretically approach this interesting regime there are
currently three options available: One is to use so called QCD
inspired quark models. Especially in spectroscopy they are widely
used and lead to very useful insights on hadron structure --- see
Ref.~\cite{quarkmodels}.  On the other hand, the application of
those approaches to hadron dynamics is technically very involved and
often only lead to a qualitative agreement with the phenomenology.
Theoretically more sound is to solve QCD numerically with large
scale simulation in a discretized space. This method is called
lattice QCD. A lot of progress was made in recent years. E.g. in
Ref.~\cite{glueballs} the glue--ball spectrum is calculated in pure
Yang--Mills theory.  In addition, first results for hadron--hadron
scattering processes at low energies were reported
recently~\cite{martin}.

The third alternative is to work directly with the physical degrees
of freedom relevant at the given energy, namely with hadrons. On the
first glance it seems as if in this approach the connection to QCD
were lost, however, with the advent of effective field theories some
30 years ago~\cite{chpt} this connection became well established. It
is nowadays possible to preform calculations with hadronic degrees
of freedom with high accuracy and contolled uncertainty. The
probalbly most impressive amongst those works is the recent
determination of the $\pi-\pi$ scattering lengths using chiral
perturbation theory (ChPT) --- the effective field theory for the
standard model at low energies --- in combination with dispersion
theory~\cite{pipi}.

The reason why it became possible to study QCD by looking at
hadronic observables directly is its approximate $SU(2)_L\times
SU(2)_R$ symmetry, spontaneously broken to $SU(2)_V$. This mechanism
leads to the appearance of the three pions as pseudo--Goldstone
bosons that are interacting only weakly at low energies. In
consequence there is, at low energies, a separation of scales
between the pion mass/momenta and the typical hadronic scale set by
$\Lambda_\chi\simeq 4\pi f_\pi \simeq M_N\simeq 1$ GeV, where
$f_\pi$ and $M_N$ denote the pion decay constant and the nucleon
mass, respectively. In addition, the structure of the interactions
of pions with each other as well as heavy fields is largely
controlled by the above mentioned symmetry group. The resulting
effective field theory is the above mentioned ChPT. For a recent
review see~\cite{ulfandveronique}.

The standard expansion parameters of ChPT are $m_\pi/\Lambda_\chi$
and $q/\Lambda_\chi$, where $m_\pi$ and $q$ denote the pion mass and
a typical momentum of the reaction, respectively.  ChPT is a
non--renormalizable field theory, however, by construction it can be
renormalized with a finite number of counter terms --- the so called
low energy constants (LECs) --- at each order in the expasion. As a
consequence with increasing orders new parameters appear in the
theory. Ideally those are to be fixed from experiment directly. In
case of a lack of experimental information or in order to understand
the phenomenological content of LECs, their values may be determined
from matching to models that contain heavier degrees of freedom
explicitly. This program, called resonance saturation, was carried
through successfully for $\pi\pi$ scattering~\cite{ressatpipi}, $\pi
N$ scattering~\cite{ulfbible}, and $NN$ scattering~\cite{ressatNN}.
In any case the predictive power of the theory emerges from the fact
the the same LECs contribute to very different reactions. Examples
of this will be given below.

%This presentation focuses on hadron induced meson production.
%First the current status of our understanding of reactions
%of the type $NN\to NN\pi$ will be presented. Then theoretical
%aspects of the production of two-pions as well as heavier
%mesons, like the $\eta$, will be dicussed.

\section{$NN\to NN\pi$}

A first step to calculate elastic and inelastic pion reactions on
the few nucleon system was taken by Weinberg already in 1992
\cite{wein}. He suggested that all that needs to be done is to
convolute transition operators, calculated perturbatively in
standard chiral perturbation theory (ChPT), with proper nuclear wave
functions to account for the
 non--perturbative character
of the few--nucleon systems. This procedure combines the distorted
wave born approximation, used routinely in phenomenological
calculations, with a systematic power counting for the production
operators. Within ChPT this idea was already applied to a large
number of reactions like  $\pi d\to \pi d$ \cite{beane}, $\gamma
d\to \pi^0 d$ \cite{kbl,krebs}, $\pi {^3}$He$\to \pi {^3}$He
\cite{baru}, $\pi^- d\to \gamma nn$ \cite{garde}, and $\gamma d\to
\pi^+ nn$ \cite{lensky,nnscatlength}, where only the most recent
references are given.

The central concept to be used in the construction of the transition
operators is that of reducibility, for it allows one to disentangle
effects of the wave functions and those from the transition
operators. As long as the operators are energy independent, the
scheme can be applied straight forwardly~\cite{wallacephillips},
however, for energy dependent interactions more care is necessary,
for typically induced non--reducible pieces appear. For the reaction
$pp\to d\pi^+$ this is demonstrated in detail in
Ref.~\cite{lensky2}, where a subtle interplay between these and
other one  loop  integrals was found.

Using standard ChPT especially means to treat the nucleon as a heavy
field. Corrections due to the finite nucleon mass, $M_N$, appear as
contact interactions on the lagrangian level that are necessarily
analytic in $M_N$.
 However,
some pion--few-nucleon diagrams employ few--body singularities that
lead to contributions non--analytic in $m_\pi/M_N$, with $m_\pi$ for
the pion mass. In Ref. \cite{recoils} it is explained how to deal
with those.

A problem was observed when the original scheme by Weinberg was
applied to the reactions $NN\to NN\pi$ \cite{park,unserd,ulfnovel}:
 Potentially higher order corrections turned out to be large and lead
to even larger disagreement between theory and experiment.  For the
reaction $pp\to pp\pi^0$ one loop diagrams that in the Weinberg
counting appear only at NNLO where evaluated \cite{dmit,ando} and
they turned out to give even larger corrections putting into
question the convergence of the whole series.  However, already
quite early the authors of Refs. \cite{bira1,rocha} stressed that an
additional new scale enters, when looking at reactions of the type
$NN\to NN\pi$, that needs to be accounted for in the power counting.
 Since the two nucleons in the initial state need to have
sufficiently high kinetic energy to put the pion in the final state
on--shell, the initial momentum needs to be larger than
\begin{equation}
p_{\rm thr} = \sqrt{M_Nm_\pi} \ \longrightarrow p_{\rm
thr}/\Lambda_\chi \simeq  \ 0.4. \label{expand}
\end{equation}
The proper way to include this scale was presented in Ref.
\cite{ch3body} and implemented in Ref. \cite{withnorbert} --- for a
recent review see Ref. \cite{report}. As a result, pion $p$-waves
are given by tree level diagrams up to NNLO in the modified power
counting and the corresponding calculations showed satisfying
agreement with the data \cite{ch3body}. However, for pion $s$--waves
loops appear already at NLO. Those were studied in detail in
Ref.~\cite{lensky2}. The loops turned out to undergo sizable
cancellations. The net effect of going to NLO was that the most
important operator (upper diagram of Fig.~\ref{pipiprod}, left
panel), first investigated in Ref.~\cite{koltunundreitan} got
enhanced by  a factor of 4/3 which was sufficient to overcome the
apparent discrepancy with the data.  Since the Delta--nucleon mass
difference, $\Delta$, is numerically of the order of $p_{thr}$, also
the Delta--isobar should be taken into account explicitly as a
dynamical degree of freedom~\cite{bira1} --- in line with
phenomenological findings~\cite{jouni,ourdelta}.

Once the reaction $NN\to d\pi$ is understood within effective field
theory one is in the position to also calculate the so--called
dispersive and absorptive corrections to the $\pi d$ scattering
length. With these results, presented in Ref.~\cite{pid1,pid2}, a
high accuracy determination of the $\pi N$ scattering lengths using
information on $a_{\pi d}$ becomes possible --- for a review of the
latter subject we refer to Ref.~\cite{pidrev}.

\begin{figure}[t]
%\psfrag{t1}{{$NN\to NN\pi$}} \psfrag{t2}{{$\pi^-d\to \gamma nn$}}
%\psfrag{t3}{{$pp\to d\nu_e e^+$}} \psfrag{t4}{{$pd\to pd$}}
%\psfrag{v}{$d(N^\dagger N)(N^\dagger \vec \sigma_k \vec \tau_i
%N)\vec \nabla_k \pi_i$}
\centerline{\psfig{file=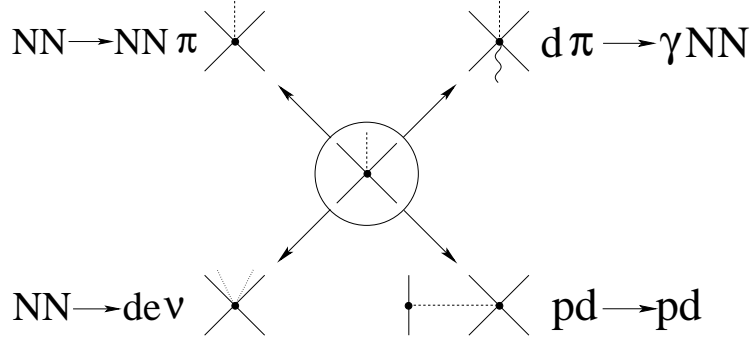,width=10cm}} \vspace*{8pt}
\caption{\label{4npi}Illustration of the various reactions, where
the leading $4N\pi$--contact term contributes.}
\end{figure}

 In Ref.~\cite{ch3body} it was stressed that there is a close connection
between the leading $(\bar NN)^2\pi$ counter term and an important
term in the three--nucleon force. The latter is discussed, e.g., in
Ref.~\cite{pdchiral}. In addition, it is again the same operator
that also appears in $\gamma d\to\pi^+NN$ and $\pi d\to \gamma
NN$~\cite{gardestig} as well as in weak reactions like tritium beta
decay~\cite{park2,phillips}. The appearance of the same operator in
various reactions is illustrated in Fig.~\ref{4npi}. Note that the
same operator appears in all these reactions in very different
kinematics ranging from very low energies for both incoming and
outgoing $NN$ pairs in $pd$ scattering and the weak interactions up
to relatively high initial energies for the $NN$ induced pion
production. In Ref.~\cite{nakamura} an aparent discrepancy between
the strength of the counter term needed in $pp\to pn\pi^+$ and in
tritium beta decay was reported. This difference might indicate a
non--applicability of ChPT to the reactions $NN\to NN\pi$. On the
other hand, in the partial wave analysis of the $pp\to
pn\pi^+$~\cite{flamang} it was assumed that the isospion 1 $NN$
final state contributes in the $S$--wave only. However, it was shown
in Refs.~\cite{complete,pwpi0} that is assumption is not justified.
Further experimental and theoretical studies are clearly called for
to resolve this issue. Here the planned double--polarization
measurements at COSY are very important~\cite{ANKE}.

As mentioned above the mechanism of spontaneous symmetry breaking
constraints the interactions of pions in a very non--trivial way.
One of the most striking examples is that the contributions to the
proton--neutron mass difference get linked to the leading, isospin
violating $\pi N$ scattering~\cite{isowein,sven}. Since it is
possible to manipulate the total isospin of a hadronic reaction
using light nuclei either in the final or the initial state, pion
production in $NN$ and $dd$ collisions proved as ideal to study
these isospin violations. The measurements of the forward--backward
assymmetry in $pn\to d\pi^0$~\cite{pndpiexp} and of the total cross
section for $dd\to\alpha\pi^0$~\cite{ddalphaexp} show clearly the
presence of isospin violation also in hadron dynamics beyond effects
resulting from pion mass differences. Theoretically first steps for
a consistent analysis of the two reactions were taken in
Ref.~\cite{pndpitheo} and Refs.~\cite{ddalphatheo}, respectively. It
is important to note that the operator structure of the leading
operators in the reactions mentioned above, which is fixed by the
assumed symmetry properties of the QCD+QED, also appears in isospin
violating decays of $D_s(2317)$ mesons~\cite{ddecay}. This again is
a good example how effective field theories by connecting different
reactions allow for deeper insights into the underlying theory. In
order to establish this connection better data for both systems is
needed. In case of $dd\to\alpha \pi^0$ this will be provided by WASA
at COSY~\cite{WASA}, in case of the $D_s$ decay by PANDA at FAIR.

\section{Two--pion production}

\begin{figure}[t]
%\psfrag{r1}{\small{$NN\to NN\pi$}} \psfrag{r2}{\small{$NN\to
%NN\pi\pi$}}
\centerline{\psfig{file=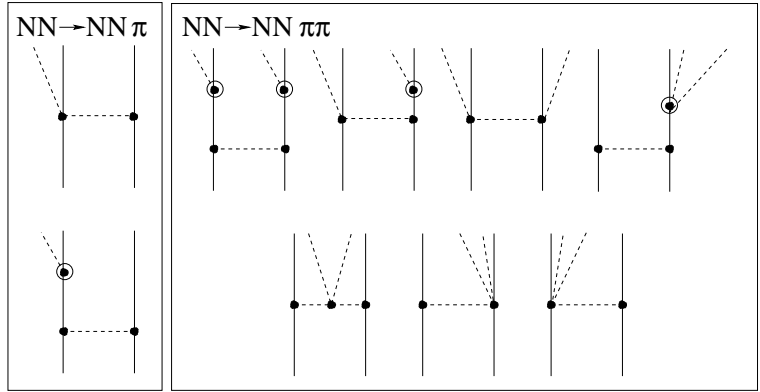,width=10cm}}
\vspace*{8pt} \caption{\label{pipiprod} Leading nucleonic diagrams
that contribute
  to $NN\to NN\pi$ (left panel) and $NN\to NN\pi\pi$ (right panel)
  near threshold.  Solid (dashed) lines denote nucleons (pions), solid
  dots (encircled dots) denote vertices from the leading
  (next--to--leading) Lagrangian. }
\end{figure}

In recent years the reactions $NN\to NN\pi\pi$ was studied both
phenomenologically~\cite{pipitheo} as well as
experimentally~\cite{pipiexp,pipiexp2,pipiexp3}. However, so far no
effective field theory calculation exists for this class of
reactions.

In case of two--pion production the threshold momentum is even
larger and accoringly the expansion parameter relevant for reactions
of the type $NN\to NN\pi\pi$ is $p_{\rm
  thr}/\Lambda_{\chi}=0.54$, which makes the applicability of ChPT
here questionable. However, it is still instructive to investigate
the structure of diagrams that contribute at the lowest orders. In
the right panel of Fig.~\ref{pipiprod} the nucleonic diagrams are
shown that contribute to leading order to two--pion production in
nucleon--nucleon collisions at threshold. Note that at leading order
for some of these there are counter parts that contain the
Delta--isobar.

For comparision in the left panel we show those that contribute to
leading order to $NN\to NN\pi$. Obviously, the number of diagrams
that contributes increased significantly. It remains to be seen how
well this set of amplitudes describes the production amplitude near
threshold. In this context it is important to note that the
phenomenological amplitudes of Ref.~\cite{pipitheo} are not able to
explain the empirical data~\cite{pipiexp2,pipiexp3}.  In this work
only some of the diagrams shown above are included.

Again, already at next--to--leading order loops start to contribute.
In the two pion production those consist of two sucessive one pion
emissions. These loops are known to be finite\footnote{The same
  class of diagrams was studied in different kinematics in
  Refs.~\cite{pid1,pid2}.} and do not introduce any new parameters. At
N$^2$LO, additional loops contribute. Those are divergent and,
consequently, at this order the first counter--terms enter the
calculation. Note, the concept of resonance saturation allows one to
identify these counter terms with some of the baryon resonance
contributions that were included in the calculations of
Ref.~\cite{pipitheo}.  In calculations for single pion production in
$NN$ collisions the Delta isobar needs to be kept as a dynamical
degree of freedom. Tree level diagrams including the Delta therefore
appear already at rather low orders. It remains to be seen, if a
similar promotion of resonace diagrams is necessary for additional
resonances in case of the two pion production. A good candidate for
this might be the Roper resonance that is believed to have a
significant two pion contribution~\cite{oli} which results in a
sizable empirical coupling to the two--pion
channels~\cite{sonja,pipitheo,zou}.

Nowadays a large number of differential and even polarized data are
available for the two pion reactions in nucleon--nucleon
collisions~\cite{pipiexp3}. It will be very important to investigate
the interplay of the diagrams controlled by chiral symmetry (see,
e.g., Fig.~\ref{pipiprod}) and the resonance contributions as the
excess energy increases. Especially to analyse the polarization data
a proper inclusion of final and initial state interactions is
essential, for analysing powers are given by the imaginary parts of
interference terms and the $NN$--distorsions are the essential
source of non--vanishing phases. These issues will be investigated
in the coming year.

\section{Production of heavier mesons}

Was it already questionable, if the chiral expansion can still be
applied for the two pion production processes, it is out of question
that there is no way to use it to study the production of heavier
mesons. Already for $\eta$ production, we get $ p_{\rm
  thr}/\Lambda_\chi \simeq \sqrt{m_\eta/M_N} = 0.8 \ .  $ Does this
mean that there is no model independent approach possible in this
case? Fortunately the answer to this question is no. It is well
known that for large momentum transfer reactions the final state
interactions become universal~\cite{goldbergerwatson}. Thus, as soon
as one studies a reaction with sufficiently strong final state
interactions, it is possible to disentangle those from the
production operator in a model--intependent fashion. The signal of
the final state interaction is a distorsion of the invariant mass
distributions of outgoing particles. It should be stressed that it
is non--trivial to make the connection of the signal of final--state
interactions to the scattering parameters quantitative~\cite{achot}
(for elastic final state interactions a method derived in
Ref.~\cite{achot2}) and that in the presence of such strong final
state interactions a determination of the strength of the total
production amplitude seems not to be possible model
independently~\cite{mitkanzo}.

A very nice example for the given discussion is the observation of
the strong $\eta{}^3$He interaction in the reactions $\gamma
^3$He$\to\eta{}^3$He~\cite{pfeiffer} and $pd\to
\eta{}^3$He~\cite{saclay,heetaanke,heetacosy11} -- see also talk by
A. Khoukaz at this conference.  In both reactions a very pronounced
rise in the amplitude is seen very close to the production threshold
pointing at a very near--by singularity in the $\eta{}^3$He
scattering amplitude~\cite{etaanal}. In addition, the angular
assymmetry shows a very unusal energy
dependence~\cite{heetaangeldep}. In order to decide whether the
$\eta$--He interaction is sufficiently strong to form a bound state,
additional sub--threshold data is necessary. Such data
exists~\cite{pfeiffer}, however, not yet of sufficient accuracy to
decide on the subject~\cite{etacomment}.

In some cases it is also possible to deduce information on the
production operator itself directly from observables even in the
case of the production of heavy mesons.  A nice example is
vector--meson production in nucleon--nucleon collisions. There was
the hope that significant deviations from predictions based on the
OZI--rule~\cite{ozi} in case of strangeness production might allow
one to draw conclusions on the strangeness content of the
nucleon~\cite{ellis,titov}.  The OZI ratio was also studied in
nucleon--nucleon induced reactions via a comparison of $\phi$ and
$\omega$ production cross sections.  Clearly, although the absolute
value of the matrix elements can at present not be controlled
theoretically, the ratio of production cross sections contains
valuable information that my be extracted from the data using the
method explained in  Ref.~\cite{phipenta} The angular distributions
of the $\phi$ production near threshold revealed that the reaction
is dominated by a meson exchange current and not by vector meson
emission off a single nucleon~\cite{omegamitkanzo,phimitkanzo} and
thus no direct information on the strangeness content of the nucleon
can be extracted.  On the other hand is a deviation of the
expectations of the OZI rule even in this case interesting. In this
context it is important to understand the role of resonances in the
vector meson production --- see, e.g. Refs.~\cite{titov2,kanzoneu}.
Although a recent experimental study indicates that the reaction
mechanisms might be the same for $\omega$ and $\phi$ production in
$NN$ collisions~\cite{omegaphi}, further studies, especially with
polarized beam, are necessary, before this issue can be resolved.
Especially additional information on differential observables might
reveal the role played by a possible pentaquark that strongly
couples to $\phi p$~\cite{phipenta}.

\section{Outlook}

%In this presentation an overview was presented over status and
%potential of meson production in nucleon--nucleon collisions. Topics
%included chiral perturbation theory, isospin violation, single and
%double pion production. In addition, interesting effects in the
%production of heavier mesons were discussed.

In the years to come significant theoretical progress in the pion
production reactions is to be expected.  Single pion production will
be caluclated up to N$^2$LO.  In addition two-- and three--pion
production on the two--nucleon system and pion production on
few--nucleon systems will be studied.  This will pave the way for
high accuracy calculations for the isospin violating reactions. For
the production of heavier mesons better data will provide deeper
insight, e.g., into the interactions of unstable particles.

To summarize, in recent years in the field of non--strange meson
production in hadronic collisions  significant progress was made on
both the theoretical as well as the experimental side. The years to
come promise deep insights into the structure and dynamics of
strongly interacting particles.

\section*{Acknowledgements}

I would like to thank   V.~Baru, J.~Haidenbauer,  A.~E.~Kudryavtsev,
V.~Lensky, and U.-G.~Mei\ss ner for a very educating and productive
collaboration.

\end{document}